# Building Change Detection Based on Multi-scale Filtering and Grid Partition

BI Qi, QIN Kun*, ZHANG Han
School of Remote Sensing and Information Engineering
Wuhan University
Wuhan, P. R. China
*Corresponding author, e-mail: qink@whu.edu.cn

HAN Wenjun
State Grid Economic and Technical Research Institute
Co. Ltd.
Beijing, P.R. China

LI Zhili, XU Kai
Faculty of Information Engineering
China University of Geosciences
Wuhan, P.R. China

*Abstract*—Building change detection is of great significance in high resolution remote sensing applications. Multi-index learning, one of the state-of-the-art building change detection methods, still has drawbacks like incapability to find change types directly and heavy computation consumption of MBI. In this paper, a two-stage building change detection method is proposed to address these problems. In the first stage, a multi-scale filtering building index (MFBI) is calculated to detect building areas in each temporal with fast speed and moderate accuracy. In the second stage, images and the corresponding building maps are partitioned into grids. In each grid, the ratio of building areas in time T2 and time T1 is calculated. Each grid is classified into one of the three change patterns, i.e., significantly increase, significantly decrease and approximately unchanged. Exhaustive experiments indicate that the proposed method can detect building change types directly and outperform the current multi-index learning method.

## I. Introduction

Building change detection plays a significant role in urban planning, geo-database updating, military surveillance and emergency response. From the perspective of change detection, these methods can be divided into two categories. One is to find change types (From-to) and the other is to detect changes as a binary result [1-2]. To further analyze change information, methods of the first categories are more promising but several problems like how to detect changes in semantic automatically, how to train a model with few samples and how to overcome noise in high resolution imagery still remain [3]. Change detection based on multi-indexes is a recently developed method, with the basic idea that in each image grid, complicated urban scenes can be represented as a set of semantic indexes instead of traditional low and middle level features [4]. Utilization of morphological building index (MBI) [5] in multi-indexes [4, 6] makes it appropriate to detect building changes, but a few problems need to be solved. Firstly, although multi-indexes are utilized to measure change, it still has some gap to detect change types directly. Secondly, MBI is rather time consuming with multi-scale and multi-direction morphological operations. Thirdly, the definition and measurement of change patterns needs to be further studied for automatic change pattern detection.

To address these problems, a building change detection method based on multi-scale median filtering and gird partition is proposed in this paper.

## II. Methodology

The proposed building change detection method can be divided into two stages, as illustrated in Fig.1. In the first stage, a multi-scale filtering building index (MFBI) is proposed to detect buildings in each temporal. In the second stage, images in time T1 and T2 are partitioned into grids of the same size, the ratio of each grid's building area in time T2 and T1 is calculated, and three building change patterns are detected automatically.

### A. Building detection

Multi-scale filtering building index (MFBI) is proposed to detect buildings with fast speed and moderate accuracy. The key to extract buildings in high resolution remotely sensed imagery is to fully utilize spatial information [4, 7-10]. After the generation of enhanced image, $k$ differential images from $k+1$ feature images generated by multi-scale and multi-direction linear morphological structures are averaged to calculate MBI, with the drawback of heavy consumption and noise. Median filtering is effective to remove noise [11] and multi-scale filtering is qualified to utilize multi-scale spatial information in high resolution imagery with less computation cost [12-13]. Inspired by these, in MFBI, three differential images after four-scale median filtering with scale factor 2 are averaged instead, alleviating the aforementioned heavy computation consumption of MBI. To be specific, MFBI is calculated as follows.

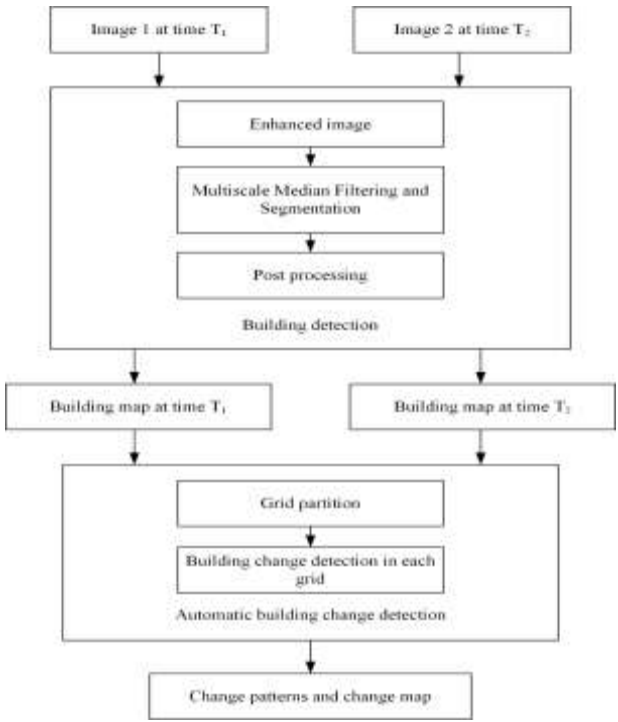

Fig. 1. Framework of the proposed method

*Step 1*. Enhanced image is generated from each pixel's maximal spectral value, the same as the implementation in [5], since buildings tend to be bright in imagery.

*Step 2*. Four-scale median filtering (with initial window size 3 and scale factor 2) on enhanced image is applied. Then, three corresponding differential images are averaged and normalized into [0, 1] to generate MFBI.

*Step 3*. Ostu method [14] is applied to segment regions belonging to buildings.

*Step 4*. NDVI and NDWI are calculated to remove some false alarms caused by vegetation and water, the same as the implementation in [15].

After the above four steps, building map of each temporal is generated.

*B. Building change detection*

The proposed method defines three building change patterns, hoping to further study the essential of change. The key of the current MBI based building change detection method [4] is that, changes in each grid are judged by the difference of the number of building pixels in time T1 and T2 with a threshold. However, it can only output a binary result (changed or unchanged), and can not find change types directly. On the other hand, it is acknowledged that in change detection, ratio method is more fault-tolerant than image differencing method. Thus, it is natural to utilize ratio strategy instead of the difference strategy in current MBI method and mainly describe building changes from the perspective of building areas as the current MBI based method does.

To realize this, we first partition images into grids of the same length. The image height and width is partitioned into N equal-length line segments respectively. In each grid, the ratio of building areas A2 in time T2 and A1 in time T1 is calculated and three change patterns, namely significantly increase, significantly decrease and approximately unchanged, are decided by a change threshold $T(T > 1)$ according to the following rules.

(1) if $A2/A1 > T$, building areas have increased significantly;

(2) if $A2/A1 < 1/T$, building areas have decreased significantly;

(3) if $1/T < A2/A1 < T$, building areas are approximately unchanged.

In current study like [4] and [6], the grid size is suggested to be set about 80 to 200 meters to cover information in urban scenes well. In this paper, N is set to satisfy the above requirements.

III. EXPERIMENTS

*A. Datasets*

Two datasets with registered and pan-sharpened bitemporal images are tested.

QB-Wuhan: Two study areas cover the east of Optics Valley in Wuhan, where great urban changes have taken place for the last ten years. Two sets of bitemporal Quick Bird (QB) images with the size of 2800*2800 pixels, acquired in August 2008 and July 2009 respectively, are shown in Fig.2 (a), (d), (g) and (j).

GF2-Ezhou: One study area with subtle urban changes covers the west of Ezhou. A set of bitemporal Gaofen-2 (GF2) images with the size of 4400*4400 pixels, acquired in January 2016 and January 2017 respectively, are shown in Fig.3 (a) and (b).

*B. Building detection results*

Since this paper focuses on building change pattern detection and tries to improve computation efficiency of MBI, we only compare the feature map and computation time of MBI and MFBI. No quantitative building detection accuracy assessment is evaluated here. For four-scale median filtering, the window sizes are 3, 6, 12 and 24 respectively.

For the first study area of QB-Wuhan dataset, MBI and MFBI feature maps are listed in Fig.2 (b), (e) and (c), (f) respectively. For the second study area of QB-Wuhan dataset, MBI and MFBI feature maps of bitemporal images are listed in Fig.2 (h), (k) and (i), (l) respectively. For GF2-Ezhou dataset, MBI and MFBI feature maps of bitemporal images are listed in

Fig.3 (c), (d) and (e), (f) respectively. As can be seen, MBI and MFBI has close capability to extract building areas, while MFBI is more capable to detect buildings in rural areas and is more easy to introduce false alarms like roads. However, as Tab.1 demonstrates, MFBI has much faster computation speed than MBI.

TABLE I. COMPUTATION TIME OF MBI AND MFBI FEATURE IMAGE ON 2 DATASETS (SECOND)

|  | QB-1 in 2008 | QB-1 in 2009 | QB-2 in 2008 | QB-2 in 2009 | GF2 in 2016 | GF2 in 2017 |
| --- | --- | --- | --- | --- | --- | --- |
| MBI | 69.00 | 69.53 | 68.37 | 68.92 | 248.47 | 197.82 |
| MFBI | 8.23 | 8.61 | 7.92 | 8.07 | 30.47 | 27.32 |

### C. Results of building change detection

For building change pattern detection, confusion matrices and overall accuracy are calculated to evaluate the classification accuracy of three patterns. To tolerate noise from building detection, T is suggested to be set more than 2, and is set to be 2.5 in all these experiments. Our method is compared with method proposed in [4], with N=14 (approximately 120meters/200 pixels for each grid) in QB-Wuhan and N=20 (approximately 220meters/220 pixels for each grid) in GF2-Ezhou.

As is shown in Fig.4 (g), (h) and (i), grids categorized to building areas significantly increasing (SI), significantly decreasing (SD) and approximately unchanged (AU) are painted in red, green and blue respectively. In comparison, the result of method in [4] is shown in Fig.4 (a), (b) and (c), and grids categorized to changed (C) and unchanged (UC) are painted in white and gray respectively. Detected change regions are labeled on images, as is demonstrated in the second and the fourth rows of Fig.4. The overall accuracy (OA) and confusion matrices on our datasets are shown in Fig.5. Change patterns in each grid are interpreted by two experts who are not involved in the experiments.

The proposed method achieves higher OA on all three study areas with less false alarms than the method in [4]. It is extremely apparent in the centered and below region of QB-Wuhan1, the left and upright region of QB-Wuhan2 and the lower right region of GF2-Ezhou that false alarms are much fewer. On the other hand, the proposed method is more capable to detect building change patterns of SI and AU rather than SD.

## IV. DISCUSSION

### A. Comparison of MBI and MFBI

MBI and MFBI are both capable of extract buildings. With multi-scale and multi-direction linear morphological structures, MBI is more robust to narrow objects like roads, but leads to heavy computation cost and long computation time. In contrast, with rectangular filtering window size, MFBI is much faster to extract building areas but is subject to these false alarms. Also, MBI is subject to built-up areas in rural areas, since buildings in village are often not regular in geometry. However, MFBI performs well on extracting these built-up areas, mainly because there is no multi-direction feature extraction in multi-scale filtering.

### B. Comparison of building change detection results

As Fig.3 and Fig.4 demonstrate, our proposed method can achieve higher accuracy than method in [4], and can detect the change pattern of buildings, which is beneficial to mine the essential of changes.

The overwhelming accuracy of the proposed method on the current MBI based method can be explained by the following reasons. Firstly, false alarms like roads are almost the same in each temporal for MFBI and thus have little influence on change detection. Secondly, in each grid, the strategy of ratio is more fault-tolerant than the difference of the number of building pixels in [4], and can reduce many false alarms. In other words, the proposed method is less sensitive to building detection results than the current method in [4]. Thirdly, in [4], the experimental data for change detection is Worldview2 imagery with eight multispectral bands, and can exclude the influence of bright soils and enhance the spectral information of building areas much more effectively.

### C. Influence of change threshold T

The key parameter of the proposed method is change threshold T. Fig.6 shows a series of overall accuracy on our dataset with different values of T (T=1.5, 2, 2.5, 3, 3.5, 4, 4.5). It can be clearly seen that with the increase of T, OA first increases significantly, then reaches its maximum, and later drops slightly. It can be explained that noise in high resolution imagery and false alarms introduced from building detection need a relatively larger T to tolerate for building change detection. But when the threshold of T becomes too large, much change information is lost.

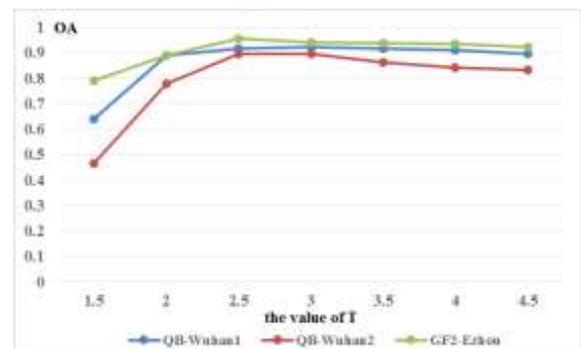

Fig. 6. Relationship between the overall accuracy and the values of T

## D. Spatial arrangement change

Except building area changes, the change of objects' spatial arrangement is also common in high resolution remotely sensed imagery. If change of spatial arrangement doesn't involve change of building areas, e.g., a circle-shaped building is located at the left and another rectangular-shaped building is located at the right in time T1 but their locations are exchanged in time T2, the proposed method is incapable to distinguish it from unchanged regions.

In [4], partitioning each grid into more cells and calculating each grid's correlation coefficient of time T1 and T2 is a partial solution to this issue but is still subject to the size of each cell.

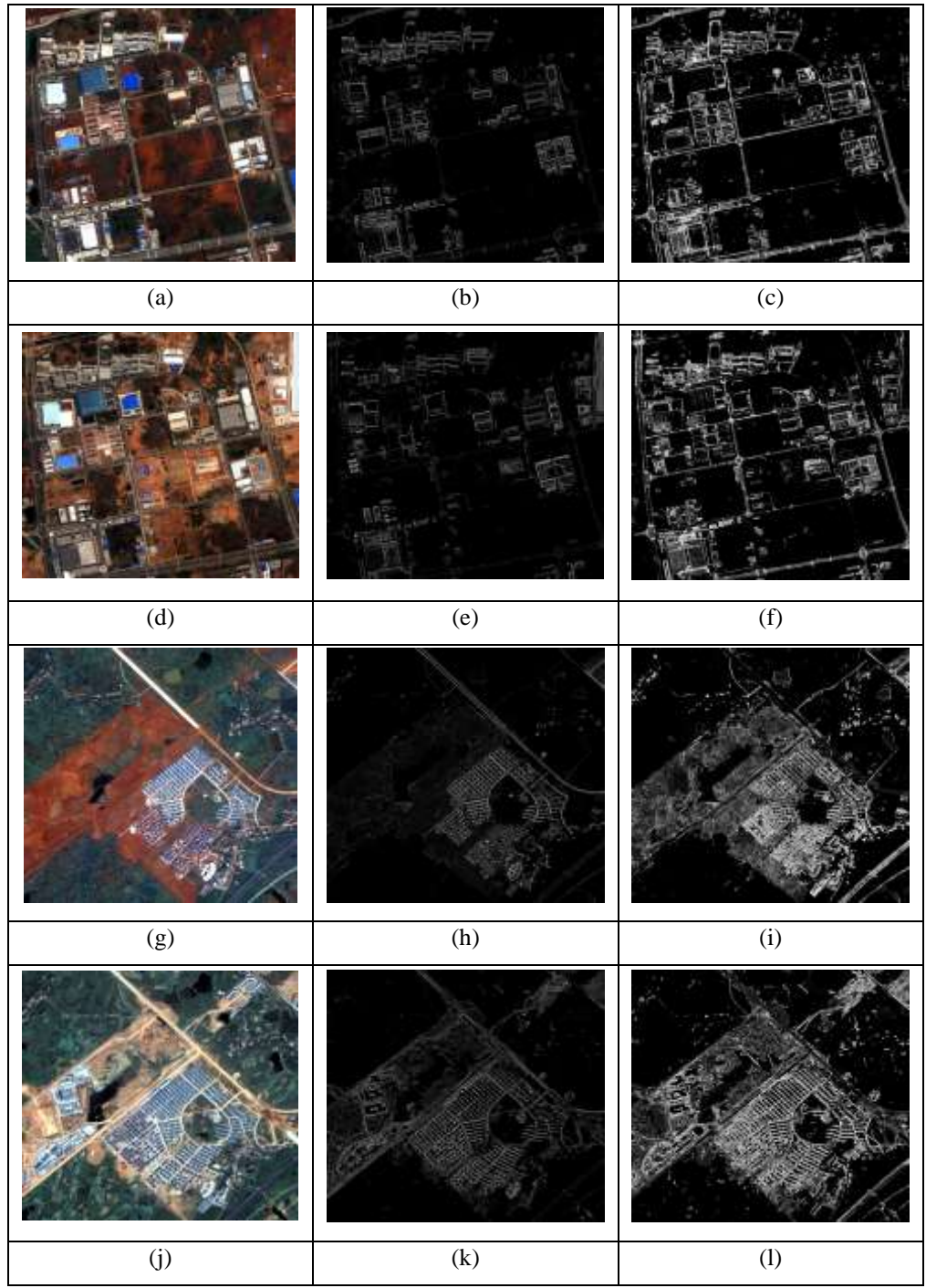

Fig. 2. (a), (b) and (c):Study area1 in 2008, its MBI and MFBI feature image; (d),(e) and (f): Study area1 in 2009, its MBI and MFBI feature image; (g),(h) and (i): Study area2 in 2008, its MBI and MFBI feature image; (j),(k) and (l): Study area2 in 2009, its MBI and MFBI feature image.

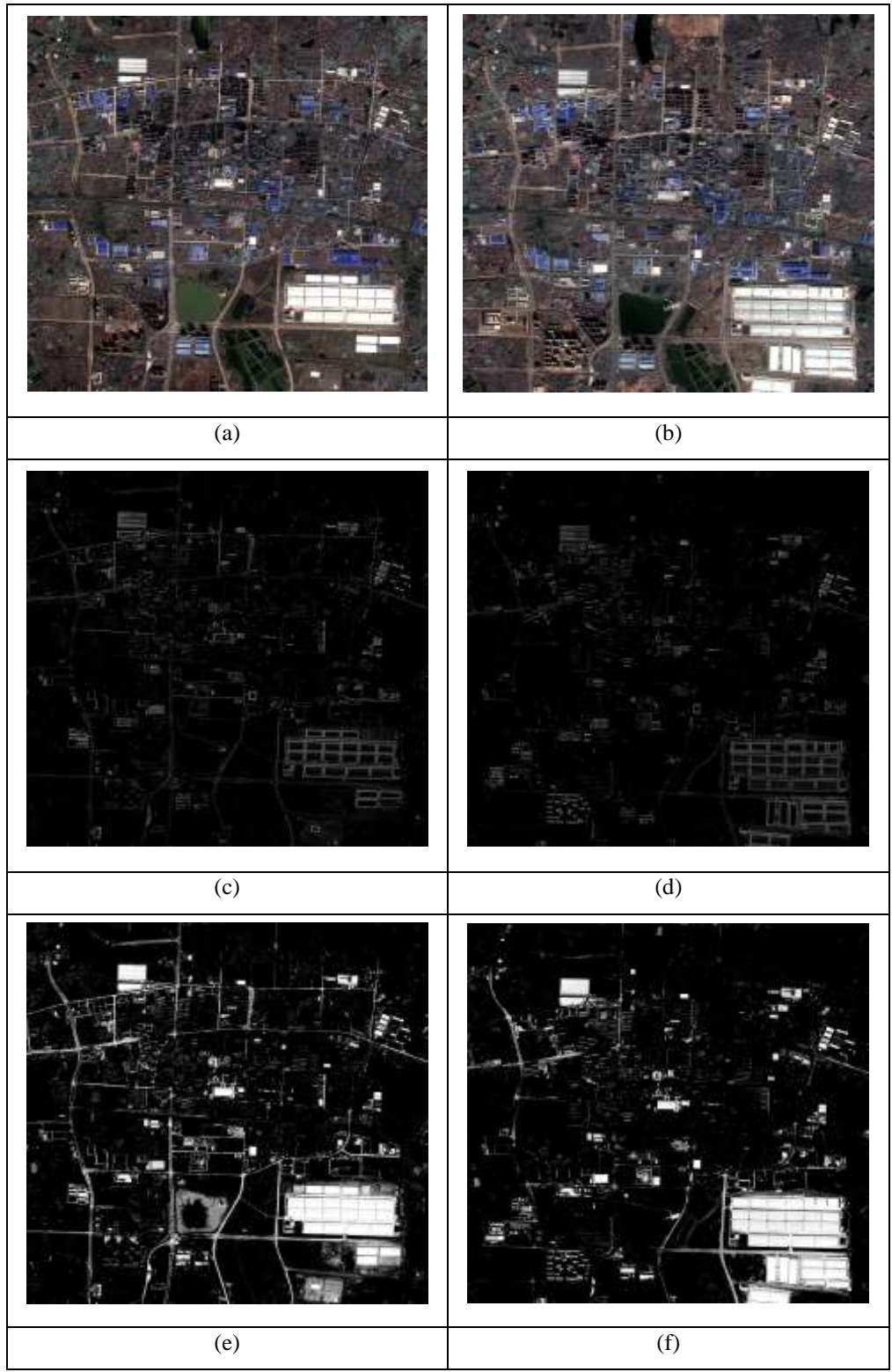

Fig. 3. GF2-Ezhou images with the corresponding MBI and MFBI feature images. (a), (c) and (e): Study area in January 2016, its MBI and MFBI feature image; (b), (d) and (f): Study area in January 2017, its MBI and MFBI feature image.

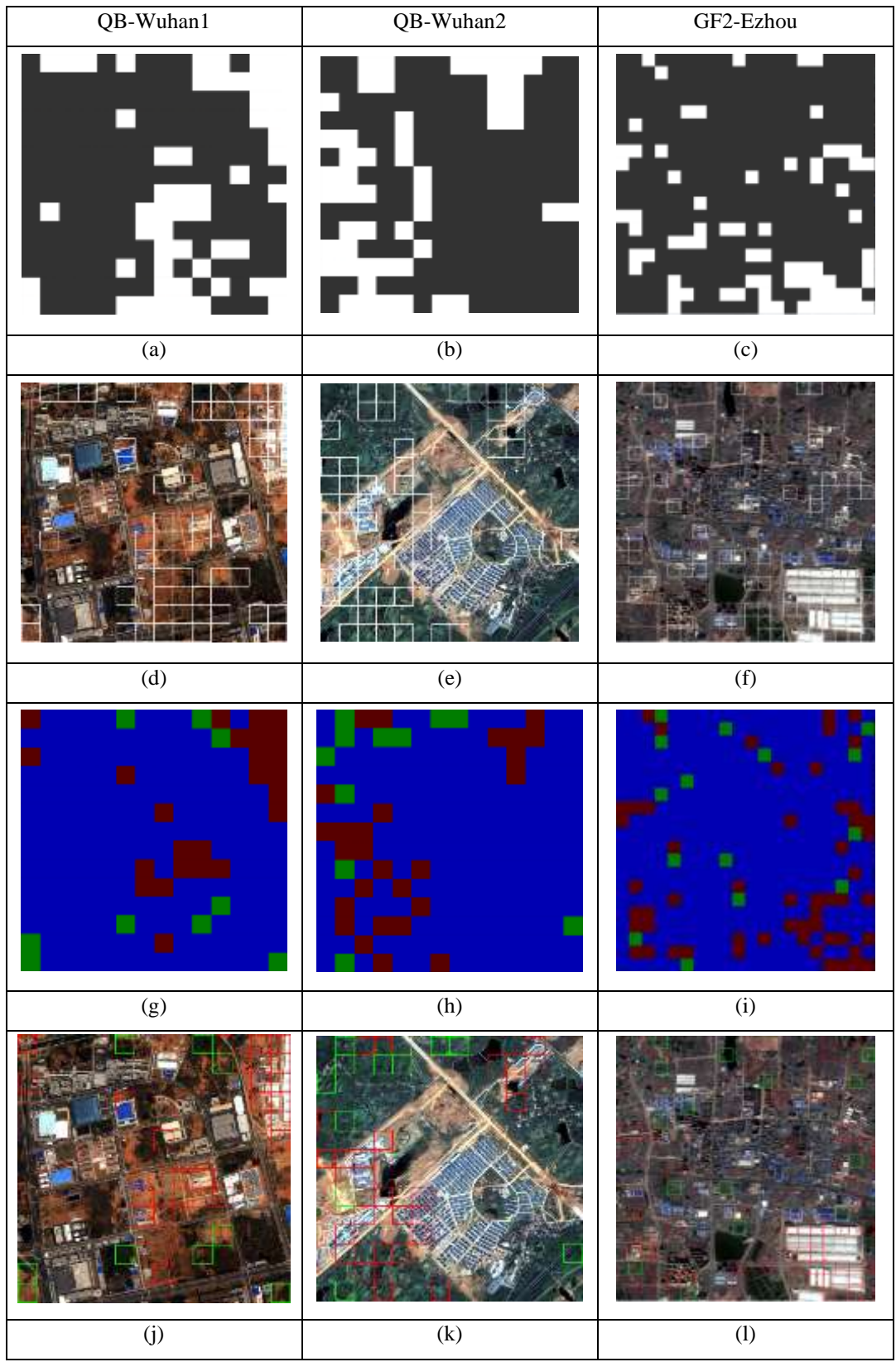

Fig. 4. First, second and the third Column: change map and change region from MBI based method and the proposed method in QB-Wuhan1, QB-wuhan2 and GF2-Ezhou, respectively.

### (a) CONFUSION MATRIX _QBWuhan1

|   | C  | UC | Tot. | %     |
|---|----|----|------|-------|
| C | 26 | 6  | 32   | 81.25 |
| UC| 28 | 84 | 112  | 75.00 |
| Tot. | 54 | 90 |   |     |
| %  | 48.15 | 93.33 | OA =76.39% | |

### (b) CONFUSION MATRIX _QBWuhan1

|    | SI | SR | AU  | Tot. | %     |
|----|----|----|-----|------|-------|
| SI | 22 | 2  | 2   | 26   | 84.62 |
| SR | 0  | 4  | 2   | 6    | 66.67 |
| AU | 3  | 3  | 106 | 112  | 94.64 |
| Tot.| 25 | 9 | 110 |    |      |
| %  | 88.00 | 44.44 | 96.36 | OA =91.67% | |

### (c) CONFUSION MATRIX _QBWuhan2

|   | C  | UC | Tot. | %     |
|---|----|----|------|-------|
| C | 26 | 3  | 29   | 89.66 |
| UC| 25 | 90 | 115  | 78.26 |
| Tot. | 51 | 93 |   |     |
| %  | 50.98 | 96.77 | OA =80.56% | |

### (d) CONFUSION MATRIX _QBWuhan2

|    | SI | SR | AU  | Tot. | %      |
|----|----|----|-----|------|--------|
| SI | 19 | 0  | 3   | 22   | 86.36  |
| SR | 0  | 7  | 0   | 7    | 100.0  |
| AU | 8  | 4  | 103 | 115  | 89.57  |
| Tot.| 27 | 11 | 106 |    |      |
| %  | 70.37 | 63.64 | 97.17 | OA =89.58% | |

### (e) CONFUSION MATRIX _GF2Ezhou

|   | C  | UC  | Tot. | %     |
|---|----|-----|------|-------|
| C | 41 | 21  | 62   | 66.13 |
| UC| 31 | 307 | 338  | 90.83 |
| Tot. | 72 | 90 |   |     |
| %  | 56.94 | 93.33 | OA =87.00% | |

### (f) CONFUSION MATRIX _GF2Ezhou

|    | SI | SR | AU  | Tot. | %     |
|----|----|----|-----|------|-------|
| SI | 46 | 0  | 2   | 48   | 95.83 |
| SR | 0  | 11 | 3   | 14   | 78.57 |
| AU | 9  | 4  | 325 | 338  | 96.15 |
| Tot.| 54 | 15 | 330 |    |      |
| %  | 85.18 | 73.33 | 98.48 | OA =95.50% | |

Fig. 5. The first ((a) and (b)), second ((c) and (d)) and third ((e) and (f)) row are confusion matrix and OA of the method in [4] and the proposed method in QB-Wuhan1, QB-Wuhan2 and GF2-Ezhou respectively.

## V. CONCLUSION

We come to the conclusion that the proposed method is effective and efficient in automatic building change detection with semantic information. Water and vegetation change detection can also be implemented in this way, only need to choose a proper index and suitable parameter values. However, the limitation lies in that building areas in each grid can't describe building changes like spatial arrangement.

Future research directions include studying parameter settings thoroughly, studying more simple and low-level feature indexes to reduce false alarms from building detection, and counting the numbers of buildings in each gird to describe changes more accurately.

## ACKNOWLEDGMENT

This research is supported by the National Key Research and Development Program of China (No. 2016YFB0502600) and the Application Research of the Remote Sensing Technology on Global Energy Internet (JYYKJXM(2017)011).